\documentclass[11pt,showpacs,preprintnumbers,amsmath,amssymb,prd,nofootinbib,superscriptaddress]{revtex4-2}

\usepackage{dcolumn}% Align table columns on decimal point
\usepackage{bm}% bold math
\usepackage{ifpdf}
\usepackage{hyperref}
\usepackage{bm}
\usepackage{xcolor,color,graphicx,graphics,physics}
\usepackage[spanish,english]{babel}%, portuguese
\usepackage[latin1]{inputenc}
\usepackage[OT1]{fontenc}
\usepackage{latexsym,amssymb,amsmath,amsfonts, slashed}
\usepackage{makeidx}
\usepackage{epsfig,subfigure}
\usepackage{natbib}
\usepackage{epstopdf}
\usepackage{mathrsfs}
\usepackage{hyperref}%\usepackage[colorlinks=true,linkcolor=blue]{hyperref}
\hypersetup{colorlinks=true, linkcolor=blue, citecolor=blue, urlcolor=blue}
\usepackage{enumerate}

\usepackage{fixmath}

\everymath{\displaystyle}
\usepackage{graphicx}

\usepackage[T1]{fontenc}
\usepackage{amsmath}
\usepackage{amssymb}
\usepackage{graphicx}
\usepackage{xcolor}

\usepackage{url}

\newcommand{\bea}{\begin{eqnarray}}
\newcommand{\eea}{\end{eqnarray}}

\newcommand{\orcid}[1]{\href{https://orcid.org/#1}{\includegraphics[width=10pt]{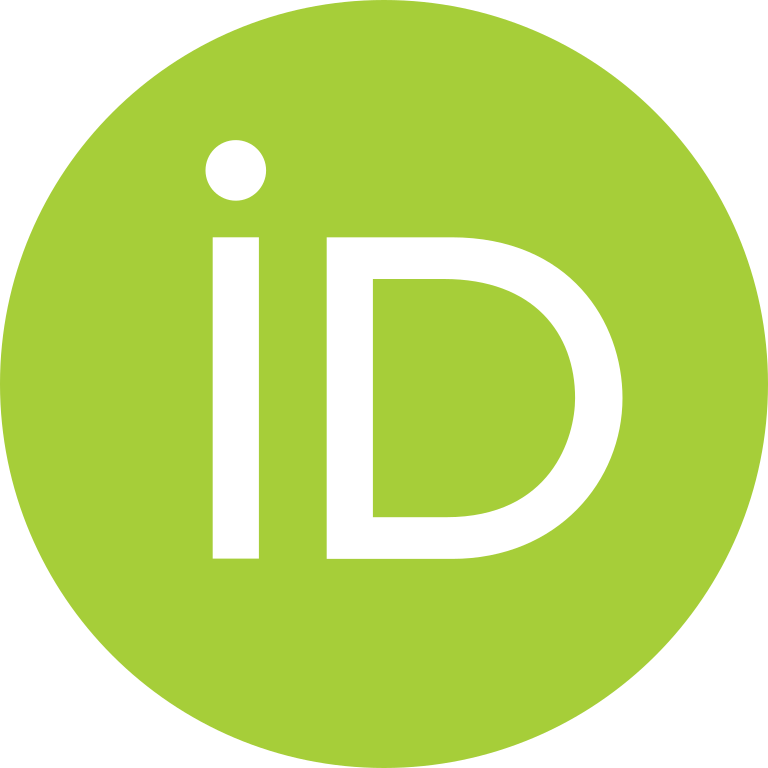}}}

\begin{document}

\title{Stefan-Boltzmann law and Casimir effect for dark photons}

\author{V. G. Prata \orcid{0009-0002-3213-601X} }
\email{victorgustavo@fisica.ufmt.br}
\affiliation{Instituto de F\'{\i}sica, Universidade Federal de Mato Grosso,\\
78060-900, Cuiab\'{a}, Mato Grosso, Brazil}

\author{A. F. Santos \orcid{0000-0002-2505-5273}}
\email{alesandroferreira@fisica.ufmt.br}
\affiliation{Instituto de F\'{\i}sica, Universidade Federal de Mato Grosso,\\
78060-900, Cuiab\'{a}, Mato Grosso, Brazil}

\author{Faqir C. Khanna \orcid{0000-0003-3917-7578} \footnote{Professor Emeritus - Physics Department, Theoretical Physics Institute, University of Alberta\\
Edmonton, Alberta, Canada}}
\email{fkhanna@ualberta.ca; khannaf@uvic.ca}
\affiliation{Department of Physics and Astronomy, University of Victoria,\\
3800 Finnerty Road, Victoria BC V8P 5C2, Canada}

\begin{abstract}

In this paper, the existence of a massive dark photon, associated with a new gauge group is considered. The dark photon can be kinetically mixed with the photon. To study some applications, the thermo field dynamics formalism is used. Exploring the topological structure of this approach, the influence of dark photons on the Stefan-Boltzmann law and the Casimir effect at zero and finite temperature is calculated.

\end{abstract}

\maketitle

\section{Introduction}

The standard model of particle physics is a gauge theory whose gauge group is~$SU(3)\times SU(2)\times U(1)$. It successfully describes three fundamental interactions, namely electromagnetic, strong and weak. Although the standard model predictions have been verified with great accuracy, it is not a fundamental theory, as it contains no gravitational interaction and does not explain the dark components of the universe, dark energy and dark matter. In the search for a solution to these problems, a physics beyond standard model emerges. In this paper, an extension of the standard model, which introduces a dark matter candidate,  is considered.

Looking specifically at Maxwell electrodynamics, an interesting extension is to include an extra Abelian boson associated with an extra $U(1)_X$ gauge group. This novel boson is called a dark or hidden photon \cite{Okun, Hol, Abel, Santiago, Fayet}. It is important to note that the visible or standard photon is taken to be the boson of the $U(1)$ gauge group of electromagnetism, while the dark photon is the boson of an extra $U(1)_X$ group. Dark photons do not directly interact with the standard model particles. However, the photon and dark photon can interact through kinetic mixing. This kinetic mixing gives a connection between the dark and visible sectors. This is a direct way to new physics beyond the standard model. In addition, it allows to detect dark photons in experiments. While this kinetic mixing leads to possible detection, a natural question is: why haven't dark matter or dark photons been observed yet? There are two possibilities: (i) the new particles may be very heavy and then a lot of energy is needed to create them. (ii) The interactions between the dark photon and those of the standard model are extremely weak, then their effects would be too weak to have been observed so far. The search for the new boson consists of astrophysical and cosmological observations, as well as laboratory experiments \cite{Book, Fil}. 
Some studies on dark photons have been developed. For a review of the main theoretical and experimental results see \cite{Jae, Fold, Marco, Cur, Den}. An interesting study carried out in the dark sector has been the calculation of the influence of a dark photon on the Casimir effect \cite{Ali}. In this paper, let's use the topological structure of the Thermo Field Dynamics (TFD) formalism to investigate the influence of dark photons on the Stefan-Boltzmann law, as well as on the Casimir effect at zero and finite temperature. The results presented here reinforce the results for the Casimir effect at zero temperature discussed in  \cite{Ali}.

TFD formalism is a thermal quantum field theory classified as a real-time approach. In this formalism, the temporal evolution can be considered together with the thermal effects. The main idea is to consider that the statistical average of an arbitrary operator is equal to its vacuum expectation value in a thermal vacuum \cite{tfd1, khannatfd, Umezawa:1982nv, Umezawa:1993yq, lifted}. In order to construct this thermal vacuum, the Hilbert space is doubled and the Bogoliubov transformation is used. The main feature of this approach is its topological structure, as it allows considering different phenomena such as the Stefan-Boltzmann law and the Casimir effect on an equal footing. In this context, any set of dimensions of the manifold can be compactified, for example, if the time dimension is compactified, temperature effects arise, while size effects emerge when a spatial dimension is compactified. Here, different topologies are considered. Then, the influence of the dark photons on the Stefan-Boltzmann law and the Casimir effect at zero and finite temperature for the electromagnetic field is investigated.

This paper is organized as follows. In section II, the dark photon theory is introduced. The energy-momentum tensor for the standard photon with influence due to dark photon is obtained. From the vacuum expectation value of the energy-momentum tensor, the propagators for both bosons, usual photon and massive dark photon, are discussed. In section III, the TFD approach is introduced and the energy-momentum tensor is written in terms of the compactification parameter. In order to obtain a finite quantity, a renormalization procedure is developed. In section IV, two applications are investigated. First, the time dimension is compactified and the Stefan-Boltzmann law with dark photons influence is calculated. Second, a spatial dimension is compactified and, as a consequence, the Casimir effect or a size effect at zero temperature is obtained. In section V, a topology that implies a combined effect of temperature and size is considered. In this case, the Casimir effect at finite temperature for the Maxwell theory with dark photons is determined. In section VI, some concluding remarks are discussed.

\section{The theory - Dark photons}

In this section, Maxwell electrodynamics is extended to include an extra Abelian boson that belongs to a new gauge group $U(1)_X$ beyond the standard model. This new boson ($X$) is electrically neutral and does not interact directly with the matter sector. The novel boson kinetically interacts with the standard model photon. Apart from mixing with the standard photon, the gauge boson remains invisible and is known as a dark or hidden photon. Therefore, it is assumed that the dark photon is a mediating particle for dark matter. 

The main objective of this section is to calculate the energy-momentum tensor associated with the new particle and then to investigate some applications using the topological structure of the TFD formalism. The Lagrangian describing the usual photon and the dark photon is given as
\begin{equation}
    \mathcal{L} = -\frac{1}{4} F_{\mu \nu} F^{\mu \nu} - \frac{1}{4} X_{\mu \nu} X^{\mu \nu} - \frac{\chi}{2} F_{\mu \nu} X^{\mu \nu} +  \frac{m^2_{\gamma}}{2} X_{\mu} X^{\mu},
    \label{lagrangianaprincipal}
\end{equation}
where $F^{\mu \nu}(x) = \partial^\mu A^\nu(x) - \partial^\nu A^\mu(x)$ is the standard field-strength tensor, $X^{\mu \nu} (x) =  \partial^\mu X^\nu(x) - \partial^\nu X^\mu(x)$ is the field-strength tensor of the dark photon, with $X_\mu$ denoting the dark $U(1)_X$ field, $\chi$ is the dimensionless kinetic mixing parameter which in some scenarios is restricted to $10^{-12}\lesssim\chi\lesssim 10^{-3}$ \cite{Roy} and $m_\gamma$ is the mass of the dark photon.

It is important to point out that the kinetic mixing term given by $\frac{\chi}{2} F_{\mu \nu} X^{\mu \nu}$ can be removed from the Lagrangian (\ref{lagrangianaprincipal}) by field re-definition. There are two ways to re-define the field and then remove the kinetic mixing term: (i) $ A_{\mu} \rightarrow  A_{\mu} - \chi X_{\mu}$ and (ii) $ X_{\mu} \rightarrow  X_{\mu} - \chi A_{\mu}$. Although these transformations are equivalent, the resulting physics is different. In the first case the dark photon becomes an uncharged massive vector particle, while in the second case, the mixture is transferred to mass terms, which implies a photon-dark photon oscillation \cite{Marco, Jae}. Here the first case is considered. Applying this re-definition, the Lagrangian (\ref{lagrangianaprincipal}) becomes
\begin{equation}
     \mathcal{L} = -\frac{1}{4} {F}_{\mu \nu}  {F}^{\mu \nu} - \frac{1}{4} {X}_{\mu \nu} {X}^{\mu \nu} + m^2_{\gamma} X_{\mu} X^{\mu}.
     \label{lagrangianafotonescuro}
\end{equation}

From the Lagrangian (\ref{lagrangianafotonescuro}), the focus is to derive the expression of the energy-momentum tensor associated with this theory. Since the Lagrangian is a function of the fields $A_\mu$ and $X_\mu$ and their derivatives, the energy-momentum tensor is defined as 
\begin{equation}
 \mathcal{T}^{\mu }_{\nu}  = \frac{\partial \mathcal{L}}{\partial(\partial_{\mu} A_{\rho})} \partial_{\nu} A_{\rho} +  \frac{\partial \mathcal{L}}{\partial(\partial_{\mu} X_{\rho})}  \partial_{\nu} X_{\rho} - \delta^{\mu}_{\nu}  \mathcal{L}.
        \label{tensorenergiamomentoparaoscampos}
\end{equation}
Applying the gauge Lagrangian (\ref{lagrangianafotonescuro}), the energy-momentum tensor is given as
\begin{equation}
     \mathcal{T}^{\mu }_{\nu}  = - F^{\mu \rho} \partial_{\nu} A_{\rho} - X^{\mu \rho} \partial_{\nu} X_{\rho} - \eta^{\mu}_{\nu} \left [ - \frac{1}{4} \left ( F_{\sigma \rho} F^{\sigma \rho} + X_{\sigma \rho}X^{\sigma \rho} \right ) + \frac{1}{2 } m^2_{\gamma} X_{\rho} X^{\rho} \right ]. 
     \label{tensorantisimetrico}
 \end{equation}
Note that this energy-momentum tensor is antisymmetric. To symmetrize it, the Belinfante-Rosenfeld method \cite{Belin, Belin2, Rosen} is used. Then a symmetric energy-momentum tensor describing the standard photon and the dark photon is obtained as
\begin{equation}
    \Theta ^{\mu }_{\nu} = -\eta^{\mu \lambda} \left (  F_{\lambda \rho} F_{\nu }^{\rho} + X_{\lambda \rho} X_{\nu}^{\rho}\right ) + \frac{1}{4} \eta^{\mu}_{\nu} \left ( F_{\sigma \rho} F^{\sigma \rho} + X_{\sigma \rho} X^{\sigma \rho} \right )   - \frac{1}{2} \eta^{\mu}_{\nu} m^{2}_{\gamma} X_{\rho}X^{\rho} .
\end{equation}
For convenience, this tensor is written as
\bea
\Theta^{ \mu \lambda}(x) &=& - F^{\mu \nu}(x) F^{\lambda }{_{\nu}}(x)  - X^{\mu \nu}(x) X^{\lambda }{_{\nu}}(x) + \frac{1}{4} \eta^{\mu \lambda} \left[    F_{\sigma \rho}(x) F^{\sigma \rho}(x) + X_{\sigma \rho}(x) X^{\sigma \rho}(x) \right]\nonumber\\
&& - \frac{1}{2}\eta^{\mu \lambda }  \eta^{\sigma \rho} m^2_{\gamma}X_{\rho}(x) X_{\sigma}(x).
\eea

In order to make some applications with this symmetric energy-momentum tensor one must calculate the vacuum expectation value of this quantity. However, this calculation is not feasible due to the presence of a product of field operators at the same point in space-time. To avoid this problem, the energy-momentum tensor is written at different points in space-time. Then
\bea
\Theta^{ \mu \lambda}(x) &=&\lim_{x'\rightarrow x}\tau  \Bigl\{ - F^{\mu \nu}(x) F^{\lambda }{_{\nu}}(x')  - X^{\mu \nu}(x) X^{\lambda }{_{\nu}}(x')\nonumber\\
&& + \frac{1}{4} \eta^{\mu \lambda} \left[    F_{\sigma \rho}(x) F^{\sigma \rho}(x') + X_{\sigma \rho}(x) X^{\sigma \rho}(x') \right]
 - \frac{1}{2}\eta^{\mu \lambda }  \eta^{\sigma \rho} m^2_{\gamma}X_{\rho}(x) X_{\sigma}(x')\Bigl\},
\eea
where $\tau$ is the ordering operator.

Assuming that both fields, i.e. $A^\mu$ and $X^\mu$, satisfy canonical quantization rules, the energy-momentum tensor becomes 
\bea
\nonumber
    \Theta^{ \mu \lambda}(x) = - \lim_{x'\rightarrow x}  \Bigl \{  \left [   \Delta^{\mu \lambda, \sigma \rho} (x,x') \left( \tau [ A_{\sigma}(x) A_{ \rho}(x') ]  + \tau [ X_{\sigma}(x) X_{ \rho}(x') ]\right)\right]\\
+4i (n^{\mu}_{0} n^{\lambda}_{0} -\frac{1}{4}\eta^{\mu \lambda})\delta(x-x') -   \frac{1}{2}\eta^{\mu \lambda}  \eta^{\rho \sigma} m^2_{\gamma} \tau[X_{\sigma}(x) X_{\rho}(x') ] \Bigl \},
\eea
where $n^\mu_0= (1, 0, 0, 0)$ is a time-like vector and 
\bea
\Delta ^{\mu \lambda, \sigma \rho} = \Gamma^{\mu\nu,\lambda}{_{\nu}}{^{,\sigma \rho}} - \frac{1}{4}\eta^{\mu\lambda}  \Gamma^{\nu \rho,}{_{\nu \rho}}{^{,\sigma \rho}}
\eea
with
\bea
 \Gamma^{\mu \nu ,{\lambda \epsilon , { \sigma \rho } }}(x,x') = ( \eta^{\nu \sigma} \partial^\mu - \eta^{\mu\sigma} \partial^{\nu}) (\eta^{\epsilon \rho} \partial^{'\lambda} - \eta^{\lambda \rho} \partial^{'\epsilon}).
\eea

In order to investigate some applications using the TFD formalism, it is important to calculate the vacuum expectation value of $\Theta^{ \mu \lambda}(x)$ which is given as
\bea
\langle \Theta^{ \mu \lambda}(x) \rangle &=& \bra{0}\Theta^{ \mu \lambda}(x) \ket{0}\nonumber\\
&=&- \lim_{x'\rightarrow x}  \Bigl \{  \left [   \Delta^{\mu \lambda, \sigma \rho} (x,x') \left( \bra{0}\tau [ A_{\sigma}(x) A_{ \rho}(x') ]\ket{0}  + \bra{0}\tau [ X_{\sigma}(x) X_{ \rho}(x') ]\ket{0}\right)\right]\nonumber\\
&&+4i (n^{\mu}_{0} n^{\lambda}_{0} -\frac{1}{4}\eta^{\mu \lambda})\delta(x-x') -   \frac{1}{2}\eta^{\mu \lambda}  \eta^{\rho \sigma} m^2_{\gamma} \bra{0}\tau[X_{\sigma}(x) X_{\rho}(x') ]\ket{0} \Bigl \}.\label{VEV}
\eea
Using the standard definition of the photon propagator, i.e.,
\bea
\bra{0}\tau[A_{\sigma}(x)A_{\rho}(x')]\ket{0} &=& i\eta_{\sigma \rho} G_{0}(x-x'), 
\eea
where 
\bea
G_{0}(x-x') &=& \frac{1}{4 \pi^2 i} \frac{1}{(x-x')^2-i \epsilon}
\eea
is the massless scalar field propagator and the dark photon propagator \cite{Prop, Prop2, Greiner} given as
\bea
\bra{0}\tau[X_{\sigma}(x)X_{\rho}(x')]\ket{0} &=& -i \left ( \eta_{\sigma \rho}+ \frac{1}{m_{\gamma}^2} \partial_{\sigma} \partial_{\rho} \right ) \Delta (x-x') 
\eea
with
\bea
 \Delta(x-x') &=& \left ( - \frac{i m_\gamma }{4 \pi^2} \frac{K_{1}(m_\gamma \sqrt{-(x-x')^2 } )}{\sqrt{-(x-x')^2 }}  \right ).
\eea
Then Eq. (\ref{VEV}) becomes
\bea
\langle \Theta^{ \mu \lambda}(x) \rangle &=&-i \lim_{x'\rightarrow x}  \Bigl \{   \Gamma^{\mu \lambda} G_{0}(x-x')  + \Sigma^{\mu\lambda}\Delta (x-x')\nonumber\\
&&+4 (n^{\mu}_{0} n^{\lambda}_{0} -\frac{1}{4}\eta^{\mu \lambda})\delta(x-x') +   \frac{1}{2}\eta^{\mu \lambda}  \eta^{\rho \sigma} m^2_{\gamma} M_{\sigma\rho}\Delta(x-x') \Bigl \},\label{VEV2}
\eea
where $\Sigma^{\mu\lambda}=-\Delta^{\mu \lambda, \sigma \rho}M_{\sigma\rho}$ with $M_{\sigma\rho}=\left ( \eta_{\sigma \rho}+ m_{\gamma}^{-2} \partial_{\sigma} \partial_{\rho} \right )$. 

This is the vacuum expectation value of the energy-momentum tensor that displays the contributions due to the visible photons plus the corrections due to the presence of the dark photons. It will be used to study some applications that emerge from the topological structure of the TFD formalism.

\section{TFD formalism}

It is well known that there are two ways to introduce temperature effects into a quantum field theory: (i) the imaginary-time or Matsubara formalism \cite{Matsubara} and (ii) the real-time formalism. The latter is divided into two approaches: closed path formalism \cite{Schwinger} and Thermo Field Dynamics (TFD) formalism \cite{tfd1, khannatfd, Umezawa:1982nv, Umezawa:1993yq, lifted}. Here, an introduction to the TFD formalism is presented.

In TFD, the system is placed in contact with a thermal reservoir, reaching thermal equilibrium. As a result, the system is described by two quantum fields: the physical field at zero temperature and the thermal field.  In this approach, the Hilbert space $S$ is duplicated, forming a thermal Hilbert space $S_T$ that is defined as $S_{T}= S \otimes \tilde{S} $, where $\tilde{S}$ is the dual Hilbert space. As a consequence, a new algebra, which map tilde and non-tilde operators, is introduced, i.e.,
\begin{eqnarray}
\nonumber
    (A_{i} A_{j})^\sim &=& \widetilde{A}_{i} \widetilde{A}_{j}, \\
    \nonumber
(cA_{i} + A_{j})^\sim &=& c^* \widetilde{A}_{i} + \widetilde{A}_{j}, \\
(A_{i}^\dagger)^\sim &=& \widetilde{A}_{i}^\dagger, \\
\nonumber
(\widetilde{A}_{i})^\sim &=& -\xi A_i, \\
\nonumber
[A_i, \widetilde{A}_j] &=& 0,
\end{eqnarray}
with $c$ being an arbitrary constant, $\xi=+1 (-1)$ for fermions (bosons) and $A$ being the standard operator for double notation. In a matrix representation, the operator is written as
\begin{equation}
A^a=
\begin{pmatrix}
A_1   \\ A_2
\end{pmatrix} =
\begin{pmatrix}
A \\ \tilde{A}^\dagger
\end{pmatrix},
\end{equation}
where $a=1, 2$. In addition to tilde operators, another element is needed in the TFD formalism, the Bogoliubov transformation, which imposes a rotation between tilde and non-tilde operators.  As an example, let's consider an arbitrary operator. Then the Bogoliubov transformation leads to
\begin{equation}
\begin{pmatrix}
A(\alpha,k)  \\ \widetilde{A}^\dagger (\alpha,k)
\end{pmatrix} = U (\alpha) 
\begin{pmatrix}
A(k) \\ \tilde{A}^\dagger(k)
\end{pmatrix},
\end{equation}
where
\begin{equation}
   U(\alpha)=\begin{pmatrix}
u(\alpha) & -v(\alpha) \\ 
 \xi v(\alpha)& u(\alpha)
\end{pmatrix},
\end{equation}
is the Bogoliubov transformation with $u^{2}(\alpha) + \xi v^{2}(\alpha) = 1$. Here, $\alpha$ is the compactification parameter defined by $\alpha=(\alpha_0,\alpha_1,\cdots\alpha_{D-1})$, where $D$ are the space-time dimensions.

Now let's apply the doubled notation, TFD formalism, to the energy-momentum tensor, or more specifically to its vacuum expectation value, which describes the Maxwell  electromagnetism with corrections due to dark photons. Then Eq. (\ref{VEV2}) dependents on the $\alpha$ parameter becomes
\bea
\langle \Theta^{ \mu \lambda(ab)}(x;\alpha) \rangle &=&-i \lim_{x'\rightarrow x}  \Bigl \{   \Gamma^{\mu \lambda} G^{(ab)}_{0}(x-x';\alpha)  + \Sigma^{\mu\lambda}\Delta^{(ab)} (x-x';\alpha)\nonumber\\
&&+4 (n^{\mu}_{0} n^{\lambda}_{0} -\frac{1}{4}\eta^{\mu \lambda})\delta(x-x')\delta^{(ab)} +   \frac{1}{2}\eta^{\mu \lambda}  \eta^{\rho \sigma} m^2_{\gamma} M_{\sigma\rho}\Delta^{(ab)}(x-x';\alpha) \Bigl \},
\eea
where the Bogoliubov transformation has been used to introduce the compactification parameter in the propagator, i.e.,
\bea
G_0^{(ab)}(x-x';\alpha)=U^{-1}(\alpha)G_0^{(ab)}(x-x')U(\alpha),
\eea
and similar results for the dark photon propagator. It is interesting to note that the TFD formalism allows constructing a thermal vacuum, $|0(\alpha)\rangle=U(\alpha)|0,\tilde{0}\rangle$. Using this definition and  the Fourier transform, the propagator can be written as
\bea
G_0^{(ab)}(x-x';\alpha)&=&i\langle 0(\alpha)| \tau[\phi^a(x)\phi^b(x')]| 0(\alpha)\rangle,\nonumber\\
&=&i\int \frac{d^4k}{(2\pi)^4}e^{-ik(x-x')}G_0^{(ab)}(k;\alpha),
\eea
with $\phi^a(x)$ being the scalar field. Among components, physical quantities are described by the non-tilde variables. Thus, the physical quantity is described by the Green function
\bea
G_0^{(11)}(k;\alpha)=G_0(k)+\xi w^2(k;\alpha)[G^*_0(k)-G_0(k)],
\eea
where $G_0(k)$ is the scalar field propagator in the momentum space and $w^2(k;\alpha)$ is the generalized Bogoliubov transformation \cite{GBT} which is defined by
\bea
w^2(k;\alpha)=\sum_{s=1}^d\sum_{\lbrace\sigma_s\rbrace}2^{s-1}\sum_{l_{\sigma_1},...,l_{\sigma_s}=1}^\infty(-\xi)^{s+\sum_{r=1}^sl_{\sigma_r}}\,\exp\left[{-\sum_{j=1}^s\alpha_{\sigma_j} l_{\sigma_j} k^{\sigma_j}}\right],\label{BT}
\eea
with $d$ being the number of compactified dimensions and $\lbrace\sigma_s\rbrace$ denotes the set of all combinations with $s$ elements.

Now a renormalization procedure is fundamental, as it allows obtaining a finite expression that describes measurable physical quantities. In order to obtain a finite expression, the procedure used here consists of 
\begin{equation}
     \Upsilon ^{ \mu \lambda (ab) }(x;\alpha) \equiv\langle  \Theta^{ \mu \lambda(ab)}(x;\alpha) \rangle  -  \langle \Theta^{ \mu \lambda (ab) }(x) \rangle,
\end{equation}
which leads  to 
\bea
\Upsilon ^{ \mu \lambda (ab) }(x;\alpha) &=&-i \lim_{x'\rightarrow x}  \Bigl \{   \Gamma^{\mu \lambda} \overline{G}^{(ab)}_{0}(x-x';\alpha)  + \Sigma^{\mu\lambda}\overline{\Delta}^{(ab)} (x-x';\alpha)\nonumber\\
&&+ \frac{1}{2}\eta^{\mu \lambda}  \eta^{\rho \sigma} m^2_{\gamma} M_{\sigma\rho}\overline{\Delta}^{(ab)}(x-x';\alpha) \Bigl \},\label{TE}
\eea
where
\begin{eqnarray}
    \overline{G}^{(ab)}_{0}(x-x'; \alpha) &=& G^{(ab)} (x-x';\alpha) - G^{(ab)}_{0}(x-x'), \\
     \overline{\Delta}^{(ab)}(x-x'; \alpha) &=& \Delta^{(ab)} (x-x';\alpha) - \Delta^{(ab)}(x-x').
    \label{100}
\end{eqnarray}

In the next section, the topological structure of the TFD formalism and Eq. (\ref{TE}) are used. Then some applications are investigated.

\section{Applications}

In this section some applications are calculated and the contribution of dark photons or dark matter is discussed. From the topological structure of the TFD formalism, three different cases are considered that imply three different topologies. The first case is the topology $\Gamma_4^1=\mathbb{S}^1\times\mathbb{R}^{3}$, where $\alpha=(\beta,0,0,0)$. This topology brings temperature effects to the system. The second topology is $\Gamma_4^1$ with $\alpha=(0,0,0,i2d)$. This leads to size effects. And the last case consists of the topology $\Gamma_4^2=\mathbb{S}^1\times\mathbb{S}^1\times\mathbb{R}^{2}$ with $\alpha=(\beta,0,0,i2d)$. In this situation, the effects of temperature and size are investigated together.

\subsection{Temperature effects with dark matter contribution}

In order to obtain the effects of temperature and the contributions due to dark photons, let's consider $\alpha=(\beta,0,0,0,)$. In this case, the time-axis is compactified into $\mathbb{S}^1$, with circumference $\beta$. For this topology the generalized Bogoliubov transformation is given as
\bea
w^2(\beta)=\sum_{l_0=1}^{\infty}e^{-\beta k^0l_0}.\label{BT1}
\eea
Using this transformation, the Green functions become
\bea
\overline{G}_0(x-x';\beta)&=&2\sum_{l_0=1}^{\infty}G_0(x-x'-i\beta l_0n_0),\\\label{GF1}
\overline{\Delta}(x-x';\beta)&=&2\sum_{l_0=1}^{\infty}\Delta(x-x'-i\beta l_0n_0),
\eea
where $\overline{G}_0(x-x';\beta)=\overline{G}^{(11)}_0(x-x';\beta)$ and $n_0=(1,0,0,0)$ is a unit time vector. With these ingredients, the energy-momentum tensor given in Eq. (\ref{TE}) is written as
\bea
\Upsilon ^{ \mu \lambda (ab) }(x;\beta) &=&-2i \lim_{x'\rightarrow x} \sum_{l_0=1}^{\infty} \Bigl \{   \Gamma^{\mu \lambda} G_0(x-x'-i\beta l_0n_0)  + \Sigma^{\mu\lambda}\Delta(x-x'-i\beta l_0n_0)\nonumber\\
&&+ \frac{1}{2}\eta^{\mu \lambda}  \eta^{\rho \sigma} m^2_{\gamma} M_{\sigma\rho}\Delta(x-x'-i\beta l_0n_0) \Bigl \}.
\eea
Taking $\mu=\lambda=0$ and performing the derivatives, the energy density is given as
\bea
 \Upsilon ^{ 0 0 (11) }(\beta) = \frac{\pi^2}{15\beta^4} + \sum_{l_{0}=1}^{\infty}\frac{m_{\gamma}}{l_{0}^2\pi^2\beta^2}\left(3m_{\gamma} K_0\left(l_{0}m_{\gamma}\beta\right) + \frac{2\left(3+l_{0}^2m_{\gamma}^2\beta^2\right)}{l_{0}\beta}K_1\left(l_{0}m_{\gamma}\beta\right)\right).\label{SBL}
\eea
The first term refers to the standard Stefan-Boltzmann law, while the second term refers to the dark matter term found by considering the presence of dark photons in our universe. Dark photons have a different dependence on temperature. Considering that the mass of the dark photon is small \cite{Marco, Caputo, Reece}, Eq. (\ref{SBL}) can be expanded and becomes
\bea
\Upsilon ^{ 0 0 (11) }(T) =\frac{\pi^2}{15}\,T^4+\left(\frac{\pi^2}{15}\,T^4+\frac{m_\gamma^2}{12}\,T^2\right).\label{small}
\eea
Note that the first term is the standard Stefan-Boltzmann law associated with the massless photon and the second term is the contribution due to the massive dark photon. It is important to emphasize that at such a low mass limit, a massive dark photon contributes to a radiance with $T^4$ that follows from the Planck law and receives a $T^2$ correction. In addition, the energy density obtained for the dark photon in Eq. (\ref{SBL}) recovers the results for the massive photon calculated in references \cite{MPH, MPH1} in both limits, very low and high temperatures. At very high temperatures the contribution is given by Eq. (\ref{small}), while at very low temperatures the energy density goes to zero exponentially.

%Therefore, the dark photon does not satisfy the same Stefan-Boltzmann law as the standard photon whose energy density increases with $T^4$. This shows that, even at high temperatures, the contribution of dark photons to the energy density is very small and the effects associated with the usual photon are clearly dominant. 

\subsection{Size effects at zero temperature with dark matter contribution}

In the topological structure of the TFD formalism, to obtain the size effects, also known as the Casimir effect at zero temperature, the compactification parameter is considered as $\alpha=(0,0,0,i2d)$. With this choice the Bogoliubov transformation is given as
\bea
w^2(d)=\sum_{l_3=1}^{\infty}e^{-i2d k^3l_3}\label{BT2}
\eea
and the Green functions are written as
\bea
\overline{G}_0(x-x';d)&=&2\sum_{l_3=1}^{\infty}G_0(x-x'-2d l_3n_3)\label{GF2}\\
\overline{\Delta}(x-x';d)&=&2\sum_{l_3=1}^{\infty}\Delta(x-x'-2d l_3n_3)
\eea
with $n_3=(0,0,0,1)$. Then the energy-momentum tensor becomes
\bea
\Upsilon ^{ \mu \lambda (ab) }(x;d) &=&-2i \lim_{x'\rightarrow x} \sum_{l_3=1}^{\infty} \Bigl \{   \Gamma^{\mu \lambda} G_0(x-x'-2d l_3n_3)  + \Sigma^{\mu\lambda}\Delta(x-x'-2d l_3n_3)\nonumber\\
&&+ \frac{1}{2}\eta^{\mu \lambda}  \eta^{\rho \sigma} m^2_{\gamma} M_{\sigma\rho}\Delta(x-x'-2d l_3n_3) \Bigl \}. \label{EMT2}
\eea
For $\mu=\lambda=0$, the Casimir energy at zero temperature is 
\bea
 \Upsilon ^{00 (11) }(d) &=& -\frac{\pi^2}{720 d^4} + \sum_{l_{3}=1}^{\infty}\frac{m_{\gamma}}{4d^2l_3^2\pi^2}  \left (-m_{\gamma}K_{0}(2dl_3m_{\gamma}) + \frac{(-1+2d^2l_3^2m_{\gamma}^2)K_{1}(2dl_3m_{\gamma})}{dl_3} \right).\label{CE}
\eea
And taking $\mu=\lambda=3$, the Casimir pressure at zero temperature is found as
\bea
 \Upsilon ^{ 33 (11) }(d)  &=&-\frac{\pi^2}{240 d^4} + \sum_{l_{3}=1}^{\infty}\frac{m_{\gamma}}{4d^2l_3^2\pi^2} \left (-3m_{\gamma}K_{0}(2dlm_{\gamma}) - \frac{(3+5d^2l^2m_{\gamma}^2)K_{1}(2dlm_{\gamma})}{dl} \right ). \label{CP}
\eea
In both results, the first term represents the standard result for Casimir energy (\ref{CE}) and Casimir pressure (\ref{CP}), respectively. While the second term is the contribution due to dark photons. The Casimir force associated with the electromagnetic field, standard photons, is attractive. In order to investigate whether the effect associated with dark matter is attractive or repulsive, let's assume that the mass of the dark photon is very small. Then the Casimir energy and pressure become
\bea
\Upsilon ^{00 (11) }(d) &=&-\frac{\pi^2}{720 d^4}+\left(-\frac{\pi^2}{720 d^4}+\frac{m_\gamma^2}{16 d^2}\right),\label{limit}\\
\Upsilon ^{33 (11) }(d)&=&-\frac{\pi^2}{240 d^4}-\left(\frac{\pi^2}{240 d^4}+\frac{m_\gamma^2}{24 d^2}\right).
\eea
In these expressions, the first term refer to the standard Casimir effect, while the second terms are contributions due to dark photons. Furthermore, the Casimir force associated with dark photons is attractive, which implies that it behaves like the standard photon. In addition, our results show that dark photons can alter the Casimir effect, at least this effect is doubled at small mass limits. Therefore, this phenomenon could be a way to investigate the real presence of dark photons in the universe. In addition to this discussion, another important point that must be discussed is whether the presence of dark photons can significantly affect the Casimir force.	Our result in Eq. (\ref{limit}) is a consequence of the theory that has been considered, i.e. the Lagrangian (\ref{lagrangianafotonescuro}). In this case, two free fields, photon field and dark photon field,  are considered, then the Casimir effect is naively doubled. An analog result has been shown for two mixed scalar fields in \cite{Mixed}. However, as discussed in \cite{Ali}, for a real situation the presence of dark photons cannot significantly affect the Casimir force. This happens because dark photons have an huge penetration value. Therefore, measurement obtained from usual boundary conditions is not possible.

\section{Size and temperature effects with dark matter contribution}

Here let's consider the effects of size and temperature at the same time. For this the compactification parameter must be chosen as $\alpha=(\beta,0,0,i2d)$. In this case the double compactification consists in one being the time and the other along the coordinate $z$. This leads to the Casimir effect at finite temperature. For this $\alpha$ parameter, the generalized Bogoliubov transformation is given as
\bea
w^2(\beta,d)=\sum_{l_0=1}^\infty e^{-\beta k^0l_0}+\sum_{l_3=1}^\infty e^{-i2dk^3l_3}+2\sum_{l_0,l_3=1}^\infty e^{-\beta k^0l_0-i2dk^3l_3}.\label{BT3}
\eea
Note that the first term is associated with the Stefan-Boltzmann law, while the second term is associated with the Casimir effect at zero temperature. Here let's focus in the third term, since it corresponds to a mixture of both effects, size and temperature. The Green functions related to the third term are 
\bea
\overline{G}_0(x-x';\beta,d)&=&4\sum_{l_0,l_3=1}^\infty G_0\left(x-x'-i\beta l_0n_0-2dl_3n_3\right),\label{GF3}\\
\overline{\Delta}(x-x';\beta,d)&=&4\sum_{l_0,l_3=1}^\infty \Delta\left(x-x'-i\beta l_0n_0-2dl_3n_3\right).
\eea
With these quantities, the expression for the energy-momentum tensor is
\bea
\Upsilon ^{ \mu \lambda (ab) }(x;d) &=&-4i \lim_{x'\rightarrow x} \sum_{l_0,l_3=1}^{\infty} \Bigl \{   \Gamma^{\mu \lambda} G_0(x-x'-i\beta l_0n_0-2dl_3n_3)\nonumber\\
&&+ \Sigma^{\mu\lambda}\Delta(x-x'-i\beta l_0n_0-2dl_3n_3)\nonumber\\
&&+ \frac{1}{2}\eta^{\mu \lambda}  \eta^{\rho \sigma} m^2_{\gamma} M_{\sigma\rho}\Delta(x-x'-i\beta l_0n_0-2dl_3n_3) \Bigl \}. \label{EMT3}
\eea
After some calculations, this equation leads to the Casimir energy at finite temperature
\begin{eqnarray}
E(\beta,d) &=& \sum_{l_{0},l_{3}=1}^{\infty} 
 \Biggl\{ \frac{8(-4d^2l_3^2 + 3l_0^2\beta^2)}{\pi^2(4d^2l_3^2+l_0^2\beta^2)^3}\\ 
 &&+ \frac{2m_{\gamma}}{\pi^2(4d^2l_3^2+l_0^2\beta^2)^{5/2}} \Bigl(  m_\gamma \sqrt{4 d^2 l^2_3 + l^2_0 \beta^2 }  (-4d^2 l^2_3 + 3 l_0^2 \beta^2) K_{0} \left ( m_{\gamma} \sqrt{4 d^2 l^2_3 + l^2_0 \beta^2 }  \right ) \nonumber  \\ 
 &&+(2(-4d^2 l_3^2 + 8 d^4 l_3^4 m^2_\gamma) +3 l_0^2 \left ( 1+2 d^2l^2_3 m_{\gamma}^2 \right )\beta^2 + l_0^4 m^2_{\gamma} \beta^4 ) K_{1} \left (m_\gamma \sqrt{4 d^2 l^2_3 + l^2_0 \beta^2 }  \right ) \Bigl) \Biggl\},\nonumber
\end{eqnarray}
where $E(\beta,d)\equiv\Upsilon ^{00 (11) }(\beta;d)$, and for the Casimir pressure at finite temperature
\begin{eqnarray}
    P(\beta,d)&=&  \sum_{l_{0},l_{3}=1}^{\infty} 
\Biggl\{ \frac{8(\beta^2 l_0^2 - 12 d^2 l_3^2)}{\pi^2(\beta^2 l_0^2 + 4d^2 l_3^2)^3}\\
&&+ \frac{m_\gamma}{\pi^2 (4 d^2 l_{3}^2 + l_{0}^2 \beta^2)^{5/2}} \Bigl(2 m_{\gamma} (-12 d^2 l_{3}^2 + l_{0}^2 \beta^2) \sqrt{4 d^2 l_{3}^2 + l_{0}^2 \beta^2} K_{0}(m_{\gamma} \sqrt{4 d^2 l_{3}^2 + l_{0}^2 \beta^2}) \nonumber\\
&&- (80 d^4 l_{3}^4 m_{\gamma}^2 - 4 l_{0}^2 \beta^2 + 3 l_{0}^4 m_{\gamma}^2 \beta^4 + 16 d^2 l_{3}^2 (3 + 2 l_{0}^2 m_{\gamma}^2 \beta^2)) K_{1}(m_{\gamma} \sqrt{4 d^2 l_{3}^2 + l_{0}^2 \beta^2})\Bigl) \Biggl\}\nonumber
\end{eqnarray}
with $P(\beta,d)\equiv\Upsilon ^{33 (11) }(\beta;d)$. The first term corresponds to the Casimir effect at finite temperature for the usual photon. The other terms are contributions due to a massive dark photon. The influence of the dark photons does not change the nature of the Casimir effect, even at high temperatures.

\section{Conclusion}

An extension of the standard model consisting of a new gauge group described by $SU(3)\times SU(2)\times U(1)\times U(1)_X$ is considered. The mediator of this new force is the dark photon that can kinetically mix with the ordinary photon. Considering the extended Maxwell Lagrangian, the energy-momentum tensor is constructed. In order to analyze some applications and calculate the dark photons influence, the TFD formalism is used. This is a thermal approach known as real-time formalism, where the time evolution of a system can be studied along with temperature effects. Temperature effects are introduced due to its topological structure. In addition to the temperature effect, it is possible to choose a different topology and, as a consequence, a size effect can be investigated. Then, the TFD formalism allows the analysis of different effects on an equal footing, such as Stefan-Boltzmann law and Casimir effect. Here the Stefan-Boltzmann law with corrections due to dark photons is calculated. Assuming that the new gauge boson has a small mass, it is shown that its contribution to the energy density has the form $aT^4+bT^2$, where $a$ and $b$ are constants, while the energy density for the usual photon is $aT^4$. The second application was obtained considering a spatial compactification that provides size effects. Then, the influence of dark photons on the Casimir effect is analyzed. In the limit of small-mass, the Casimir pressure associated with dark photons is attractive. Therefore, it has the same behavior exhibited by the standard photon. For the last investigation, a double compactification is considered. As a consequence, the temperature and size effects are calculated. These results show that, even at high temperatures, contributions from dark photons do not change the nature of the Casimir effect. It is important to highlight that, although the theme is the same, the procedures developed in this work are very different from the study carried out in \cite{Ali}. Furthermore, as discussed in \cite{Roland}, the Stefan-Boltzmann law describes the radiation spectrum in thermal equilibrium. Then it is difficult to measure the effect of dark photons on this phenomenon, since dark photons (massive photons) will have a very long equilibrium time, and thus they will have little effect on the thermodynamics of a blackbody. Another important point is related to the measurement of corrections for the Casimir effect. Due to the enormous value of the penetration depth of dark photons, it will prevent ideal boundary conditions assumed in the Casimir effect.

\section*{Acknowledgments}

This work by A. F. S. is partially supported by National Council for Scientific and Technological Develo\-pment - CNPq project No. 313400/2020-2. V. G. P. thanks CAPES for financial support.

%%%%%%%%%%%%%%%%%%%%%%%%%%%%%%%%%%%%%%%%%%%%%%%%%%%%%%%%%%%%%%%%%%%%%%%%%%%%%%%%%%%%%%%%%%%%%%%%%%%%%%%%%%%%%%%%%

\global\long\def\link#1#2{\href{http://eudml.org/#1}{#2}}
 \global\long\def\doi#1#2{\href{http://dx.doi.org/#1}{#2}}
 \global\long\def\arXiv#1#2{\href{http://arxiv.org/abs/#1}{arXiv:#1 [#2]}}
 \global\long\def\arXivOld#1{\href{http://arxiv.org/abs/#1}{arXiv:#1}}

%%%%%%%%%%%%%%%%%%%%%%%%%%%%%%%%%%%%%%%%%%%%%%%%%%%%%%%%%%%%%%%%%%%%%%%%%%%%%%%%%%%%%%%%%%%%%%%%%%%%%%%%%%%

\end{document}